\documentclass{article}
\usepackage{spconf,amsmath,graphicx}
\usepackage[medium,compact]{titlesec}
\usepackage{verbatim}

\title{UFANS: U-shaped Fully-parallel Acoustic Neural Structure for Statistical Parametric Speech Synthesis with 20x faster}
\twoauthors
  {Dabiao Ma, Zhiba Su, Yuhao Lu \thanks{$^*$ indicates the corresponding authors}}
	{Turing Robot co.ltd\\
	Multi-modal Group\\
	Beijing, China}
  {Wenxuan Wang, Zhen Li $^*$}
	{Chinese University of Hongkong, Shenzhen\\
	School of Science and Engineering\\
	Shenzhen Research Institute of Big Data (SRIBD)\\
	Shenzhen, Guangdong, China}
\begin{document}
%
\maketitle
\begin{abstract}
Neural networks with Auto-regressive structures, such as Recurrent Neural Networks (RNNs), have become the most appealing structures for acoustic modeling of parametric text to speech synthesis (TTS) in recent studies. Despite the prominent capacity to capture long-term dependency, these models consist of massive sequential computations that cannot be fully parallel. In this paper, we propose a U-shaped Fully-parallel Acoustic Neural Structure (UFANS), which is a deconvolutional alternative of RNNs for Statistical Parametric Speech Synthesis (SPSS). The experiments verify that our proposed model is over {\bf{20}} times faster than RNN based acoustic model, both training and inference on GPU with comparable speech quality. Furthermore, We also investigate that how long information dependence really matters to synthesized speech quality.
\end{abstract}
\section{Introduction}
\label{sec:intro}

Text-to-Speech (TTS) is to convert a text string to speech of a specific speaker's voice. The typical parametric TTS system extracts linguistic features from raw text by front-end text analyzer, which is then feed into an acoustic neural networks for intermediate acoustic features extraction~\cite{allen1987text}. Afterwards, the obtained acoustic features are converted to waveforms by a vocoder, e.g. WORLD \cite{WORLD}, STRAIGHT \cite{STRAIGHT}, and WaveNet based vocoder \cite{WaveNet}, to obtain the final speech. 

Recurrent neural structures have achieved great success as acoustic models due to their capacity to capture long-term dependencies. Variants like Long Short-Term Memory (LSTM) \cite{LSTM}, Gated Recurrent Unit (GRU) \cite{GRU} and other RNN structures are now broadly used in text-to-speech~\cite{Merlin} with very good records. But all the recurrent structures result in a high time latency since the computation of current step depends on the completion of previous computations. Quasi-Recurrent Neural Network (QRNN) \cite{QRNN} and Simple Recurrent Unit (SRU) \cite{SRU} tries to make most of the computations independent of the previous computations and claims that it is 5-10 times faster than an optimized LSTM on GPU. But it cannot fully utilize the power of GPU since it still preserves recurrent structures. 

CNN has also been used as acoustic model in recent studies. In \cite{blaauw2017neural}, \cite{dctts} and \cite{ping2018deep}, Convolutional layers are used to model low-dimensional audio representation in an auto-regressive manner. It effectively accelerates training, but the output variable depends linearly on its own previous values, which causes high inference latency.

U-Net, which was first proposed for image segmentation~\cite{U_Net}, is a neural network designed with one contraction path and one expansive path. Within U\_Net, those pooling and up-sampling operation along the spatial dimension make the receptive field increases exponentially and high way connection allows the combination of different scales features. Inspired by the success of U\_Net in image segmentation, U-shaped models have been proposed for various acoustic applications, e.g., denoise~\cite{liu2018boosting}, audio source separation~\cite{stoller2018wave}. 

Here we propose a fully convolutional structure UFANS motivated by U-Net\cite{U_Net} that models long time dependencies while dozens of times speed-up both on training and inference period. As far as our knowledge, no fully parallel convolutional or transposed convolutional structure including U-net has ever been proposed for acoustic decoder in speech synthesis. And we also try to answer the question that how long information dependency really matters mentioned in \cite{FSMN}.

In section 2, our UFANS model is introduced in detail. In section 3, we
describe the experiments and try to explain the important features of UFANS. And we draw our conclusion in section 4.

\section {U-shaped Fully-parallel Acoustic Neural Structure (UFANS)}


\subsection {Overall structure of our UFANS model}
\begin{figure}[h]
\begin{center}
  \includegraphics[width=3.3in]{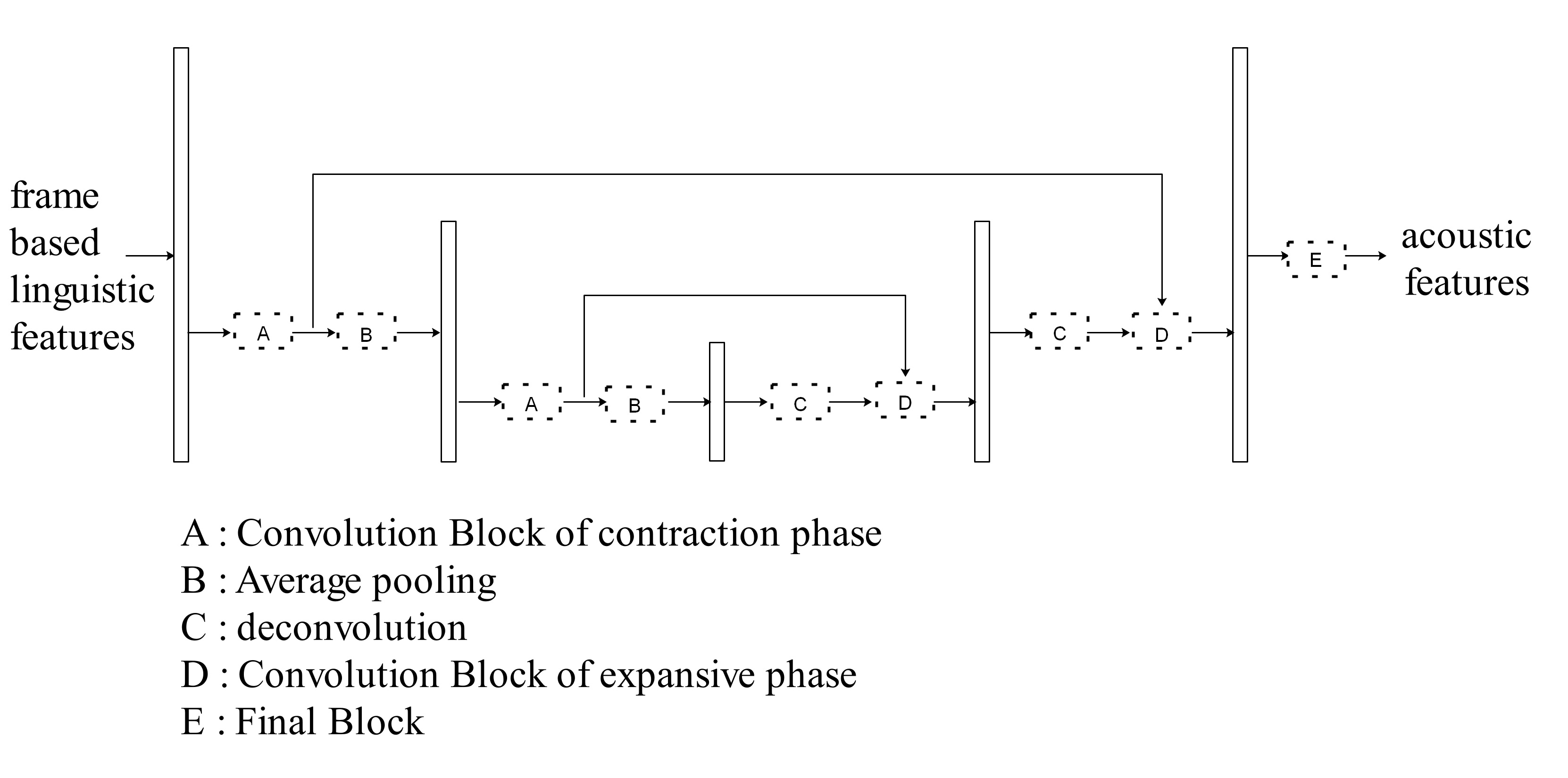}
  \caption{UFANS, with $2$ down-sampling and up-samplings.}
  \label{ufans}
\end{center}
\end{figure}

Considering the superiority of U-shaped model, we first propose an UFANS model, whose components are all parallelizable convolutional and transposed convolutional layers.The structure of UFANS model is shown in Fig.~\ref{ufans}, with 2 down-samplings and 2 up-samplings along the frame dimension. \\ \\

\begin{figure}[h]
  \includegraphics[width=3.2in]{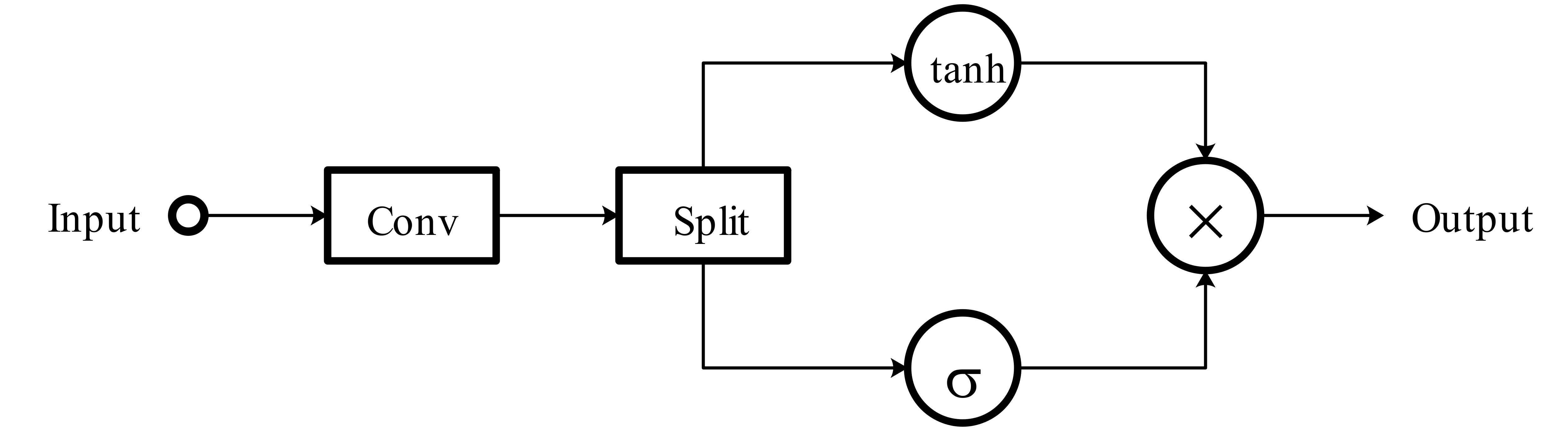}
  \caption{Convolution block (A in figure \ref{ufans} ) of contraction phase, here the operation 'split is performed on the channel size.}
  \label{conv_block}
\end{figure}

Different convolution blocks are designed sophisticated for different phases in UFANS model. Fig.~\ref{conv_block} is the revised structure of convolution block of the contraction phase in U-shaped model. Inspired by the gated activation unit used in~\cite{LSTM} and ~\cite{GRU}, we novelty split the output of convolutions layers along channel dimension in half~\cite{GatedActivation}, which enhances the capacity of UFANS to adaptively control the input flow. Besides, in UFANS, average pooling is used to keep more information in our regression task. The formula is presented as follows, \\ \\
\begin{equation}
P_1, P_2 = Split(Conv(Input))
\end{equation}
\begin{equation}
Output = tanh(P_1) * sigmoid(P_2)
\end{equation}  \label{contractionformula}

\begin{figure}[h]
  \includegraphics[width=3.2in]{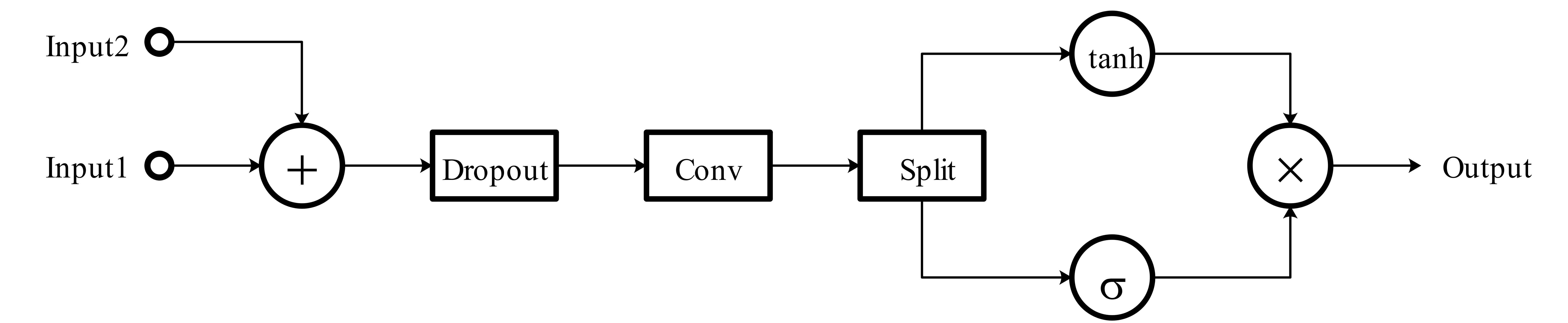}
  \caption{Convolution block (D in Figure \ref{ufans}) of expansive phase, here the operation 'split' is performed on the channel size.}
  \label{conv_block2}
\end{figure}

Correspondingly, the similar structure of convolution block in expansive phase (D in Fig.~\ref{ufans}) in presented in Fig.~\ref{conv_block2}. The input of this block consists of two parts, one from the previous layer, another from the corresponding output of the contraction phase. Besides, dropout is added here as a regulization to increase the generalization capacity of our UFANS. The formula is as follows,
\begin{equation}
P_1, P_2 = Split(Conv(Dropout(Input1 + Input2))) \\
\end{equation}
\begin{equation}
Output = tanh(P_1) * sigmoid(P_2)
\end{equation}

The final block consists of a convolution with $tanh$ activation function and a dense layer for output. It's worthy mentioning that all the components in UFANS are parallelizable convolutional layers. Thus, the model can be accelerated significantly. 


\subsection {Multi-speaker embedding}
It is straightforward to extend our UFANS model to multiple speaker cases. We insert trainable speaker embeddings into every gated activation unit to globally condition the models. These embeddings behave like biases. Now the Eqn.~(2) becomes 
\begin{equation}
Output = tanh(P_1 + h_{id}) * sigmoid(P_2 + g_{id})
\end{equation}
Where $ h_{id}, g_{id} $ are speaker embedding biases, We use different embedding biases for different layers and different phases.

\subsection {Information dependency}
UFANS can model very long information dependency since pooling make the receptive field increase exponentially. Suppose $ N $ downward and upward samplings are performed, then the number of adjacent frames that offer information to one specific output frame $ S_N $ is determined by the following iteration.
\begin{equation}
S_i = 2 * (S_{i-1} + 2), S_0 = 0, i = 1, ..., N
\end{equation}
Table 1 shows the rapid increase of information field.
\begin{table}[h!]
\begin{center}
\begin{tabular}{ |c|c|c|c|c|c|c|c| } 
 \hline
 $ N $ & 3 & 4 & 5 & 6 & 7 & 8 & 9 \\
 \hline
 $ S_N $ & 28 & 60 & 124 & 252 & 508 & 1020 & 2044 \\
 \hline
\end{tabular}
\caption{Receptive field increases exponentially}
\end{center}
\end{table}
A 5-second speech sample consist of 1000 frames, which means when $ N >= 9 $ UFANS can model information dependency of the whole utterance.

\section {Experiments}
\subsection {Dataset}
The datasets are speeches from two female Mandarin speakers, about 9000 utterances (around 10 hours,3 to 7s each) each for training, 1000 each for validation, and 500 each for test. The speech signals are sampled at 24 kHz rate, and Fourier transformed with hop length 5 ms. we extracted 60-dimensional mel-cepstral coefficients, 5-dimensional band-aperiodicity parameters, 1-dimensional logarithmic fundamental frequency and their delta, delta-delta dynamic features \cite {MLPG}, and one additional voiced/unvoiced dimension, that is 193 dimensions in total. The input features are frame based linguistic features of dimension 165, upsampled and aligned with manually annotated ground truth duration tagging. MLPG \cite {MLPG} is used to process the predicted features and WORLD \cite{WORLD} deterministic vocoder is used to synthesize waves. Both the linguistic and acoustic features are normalized before training. 
\subsection {Model hyper-parameters}
The Bi-LSTM baseline system consists of four dense layers with 1024 channels, followed by one Bi-LSTM layer with 384 channels for each direction, as is used in \cite{Merlin}. We also train a larger one that consists of one dense layer with 2048 channels followed by three Bi-LSTM layers with 1024 channels for each direction, which is used as baseline model in \cite{FSMN}. The Bi-SRU system consists of three dense layers with 1024 channels followed by four Bi-SRU layers with 512 channels for each direction. The DNN system consists of Three dense layers with 1024 channels\cite{qian2014training}.UFANS uses nine down-samplings, 3*3 kernel convolutions with 256 channels in all blocks except in the final block where a convolution with 512 channels is used. 

The frame level based mean squared error (MSE) is taken as the training criteria. We train all the models by Adam \cite{Adam} method with batch size 16, initial learning rate $ 0.0004 $ and $ \beta_1, \beta_2 $ to be $ 0.9, 0.999 $. The model is implemented with MXNET. We trained all models to achieve best validation records. Training finished in 5 hours on single Nvidia Titan X GPU. 

\begin{table*}[t]
\label{result}
\begin{center}
\begin{tabular}{ |c|c|c|c|c| } 
 \hline
  Model & MSE & Parameter size (MB) & Fully parallel & GPU Inference Time (ms) \\
 \hline
  Bi-LSTM \cite {FSMN}  & 192.8 & 292 & No & 243 \\
 \hline
  Bi-LSTM \cite{Merlin} & 195.3 & 30 & No & 69 \\
 \hline
  Bi-SRU & 194.9 & 74 & No & 44 \\
 \hline 
  UFANS & 194.6 & 42 & Yes & 3.2 \\
   \hline 
  DNN\cite{qian2014training} & 209.2 & 9.5 & Yes & 0.84 \\
 \hline
\end{tabular}
\caption{Comparison of objective results.}
\end{center}
\end{table*}
\subsection {Evalution}
Two types evaluations are applied for the performance of our generated speech, i.e., quantitative results and user study. 
\subsubsection {Quantitative results}

The quantitative results of our experiments are presented in Table 2, where MSE term is averaged between the two speaker test cases. The inference speed is evaluated as the time latency to synthesize one-second speech, which includes data transfer from main memory to GPU global memory, GPU calculations and data transfer back to main memory. The larger Bi-LSTM system got the lowest MSE while at the cost of a much larger parameter size and very low inference speed. Actually, as we will see in the following section, the speeches synthesized from UFANS even have better quality than the speeches synthesized from the larger Bi-LSTM system. Bi-SRU got slightly better MSE than Bi-LSTM system used in \cite{Merlin}, while Bi-SRU is faster. UFANS also got better MSE than Bi-LSTM in \cite{Merlin} with a speed-up factor of 23, and a speed-up of 76 compared to the larger Bi-LSTM. We did all the experiments on GeForce GTX TITAN X devices.

\subsubsection {User study}
We did a user study to compare the quality of synthesized speeches from all the systems. We randomly selected 20 Mandarin reviewers (10 female and 10 male) to listen to 20 utterances generated by each system and score each utterance. The text of the 20 utterances are randomly chosen from websites. The reviewers do not know which model the utterances come from. Results of mean opinion score (MOS) are shown in Fig.~5, where A, B, C, D represents DNN in \cite{qian2014training},Bi-LSTM in \cite {Merlin}, Bi-SRU\cite{SRU} , larger Bi-LSTM in \cite{FSMN}, and UFANS, correspondingly. A same Bi-LSTM based duration model trained on the same dataset is used in user study. 

For DNN, Bi-SRU and Bi-LSTM , the subjective MOS matches the objective evaluation. While UFANS surprisingly reaches the highest MOS with a worse objective evaluation than the larger Bi-LSTM. The reason for this behavior is to be further studied, but we believe it is related to specific features of UFANS.
 
\begin{figure}[h]
\begin{center}
  \includegraphics[height=2.5in]{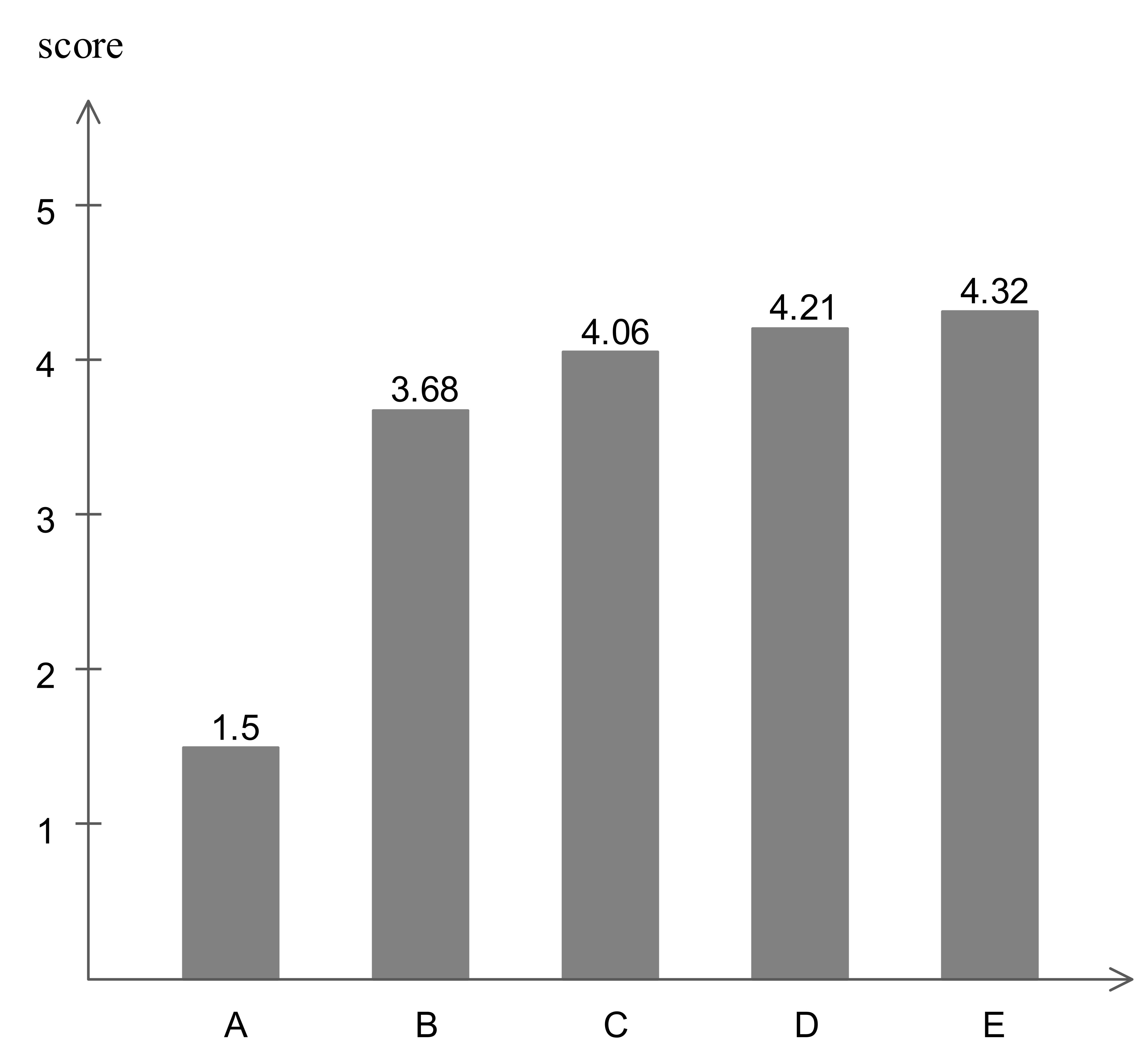}
  \caption{User study of different systems. A: DNN, B: Bi-LSTM, C: Bi-SRU, D:Large Bi-LSTM, E: UFANS.}
\end{center}
\end{figure}

\subsection {Importance of information dependency}
\begin{figure}[h]
\begin{center}
  \includegraphics[height=2.2in]{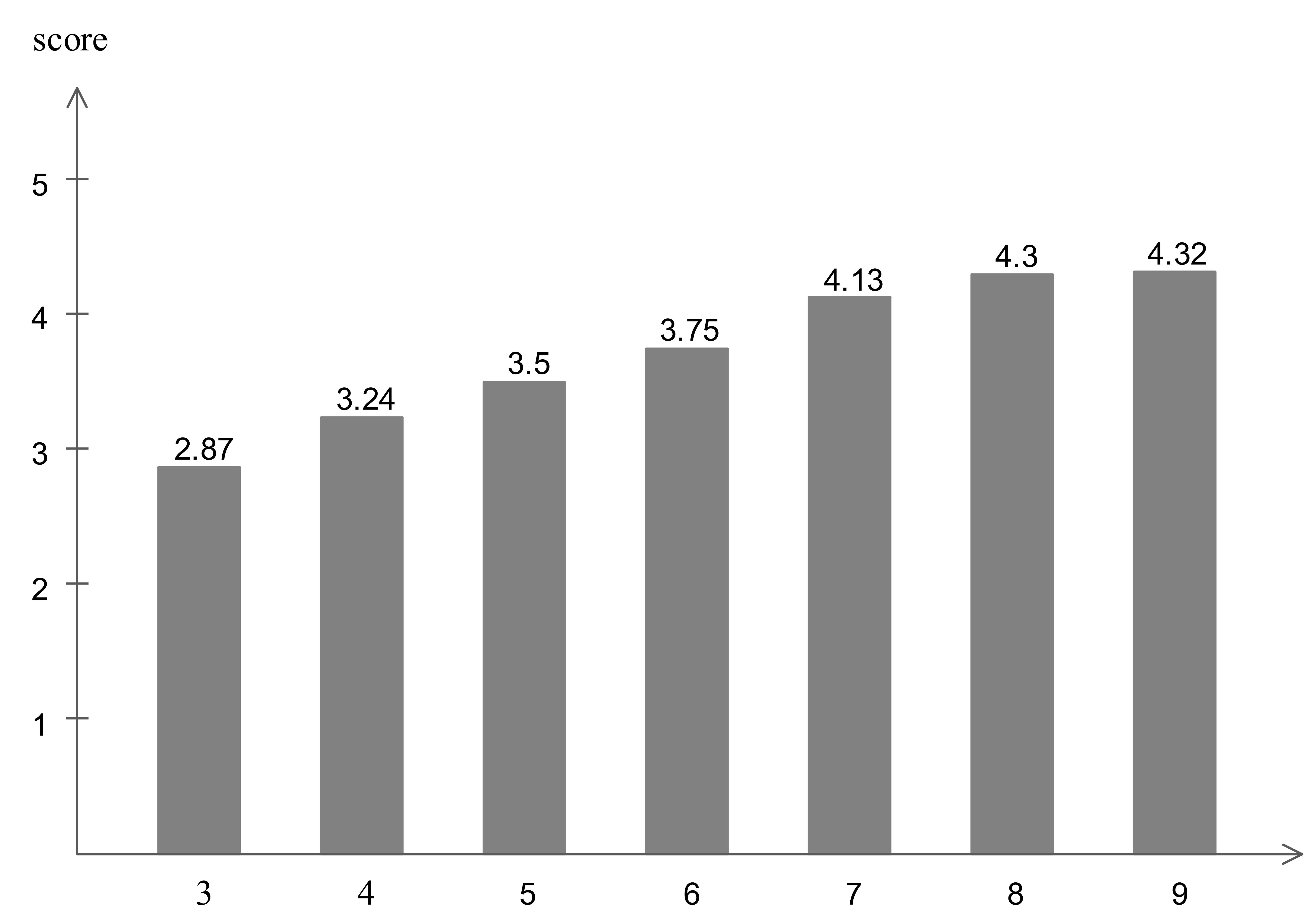}
  \caption{MOS when $ N $ ranges from 3 to 9.}
  \label{mos39}
\end{center}
\end{figure}

To answer the question that dependency within how long time really matters to synthesized speech quality, we increase the down-sampling number $ N $ of UFANS from 3 to 9 and observe how evaluation results change. Table 3 and Figure \ref{mos39} show the results.

The average frame length of utterances used in the subjective evaluation is about 1208, which means when $ N = 9 $ the information dependency covers nearly all the frames, when $ N = 8 $ the information dependency covers about half of the utterance. The MOS of $ N = 8 $ is very close to that of $ N = 9 $, and MOS begins to fall dramatically when $ N < 8 $. From this observation we can conclude that longer information dependency leads to better synthesized speech quality and dependency as long as half of the utterance is a necessity to achieve good speech quality.
\begin{table}[h!]
\begin{center}
\begin{tabular}{ |c|c|c| } 
 \hline
 $ N $ & Frame dependence & MSE\\
 \hline
 3 & 28 & 202.58 \\
 \hline
 4 & 60 & 200.84 \\
 \hline
 5 & 124 & 199.15 \\
 \hline
 6 & 252 & 197.55 \\
 \hline
 7 & 508 & 196.05 \\
 \hline
 8 & 1020 & 194.63 \\
 \hline
 9 & 2044 & 194.60 \\
 \hline
\end{tabular}
\caption{MSE when $ N $ ranges from 3 to 9.}
\end{center}
\end{table}

\subsection {Ablation study}
The fully convolutional structure of UFANS makes it possible to fully utilize the advantage of a GPU device; The rapidly increasing receptive field helps to model very long information dependency. While another important structure of UFANS is the skip connection between block A and block D (Figure \ref{ufans}). 
On one hand, In \cite {BLACK} the information bottleneck theory is applied to analyze the dynamic process of neural networks. They show that the mutual information between layers and inputs become smaller when layers go deeper, which means that information becomes more compressed through the network. The direct connection between contraction phase and expansive phase is actually a combination of higher level of abstraction and lower level of abstraction. On the other hand, the information from contraction phase focuses on local fields, while the information from expansive phase are abstracted from long fields, which means the skip connection can also be taken as a combination of local and global field information. To illustrate the importance of this connection, we remove the skip connection from our UFANS ($ N $ = 9) and find that we can only get a final MSE of about 214.5 (Table 4) which proves the importance of the connection.

\begin{table}
\begin{center}
\begin{tabular}{|c|c|c|}
 \hline
 skip connection & $ N $ & MSE \\
 \hline
 Yes & 9 & 194.60 \\
 \hline
  No & 9 & 214.51 \\
 \hline
\end{tabular}
\caption{MSE with and without skip connection.}
\end{center}
\end{table}

\section {Conclusion and Future Work}
In this paper, we propose an U-shaped Fast Acoustic Neural Structure
(UFANS). Our structure  greatly reduces time latency in the context
of GPU computation with comparable  speech quality. We show the possibility and prospect of applying fully-parallel CNN structure in TTS tasks. We
also show that the huge receptive field (long-time information dependency) and the high-way skip connection structure(combination of different level features) in our UFANS ensure the speech quality .In future, we
plan to search better structures on the basis of UFANS to get
improved synthesized speech quality. We are doing some experiments about incorporating UFANS into end-to-end TTS
system by replacing the internal RNN structures to speed up
the whole system.
\\
\\
\\
\\
\\

\bibliographystyle{IEEEbib}
\bibliography{mybib}

\end{document}